\documentclass[12pt]{article}

\newcommand {\slsh} [1] {\not{\hbox{\kern-2pt${#1}$}}}

\usepackage[latin1]{inputenc}
\usepackage[T1]{fontenc}
\usepackage[english]{babel}
\usepackage{amsmath}
\usepackage{amsfonts}
\usepackage{latexsym}
\usepackage{indentfirst}
\usepackage{epsfig}
\usepackage{graphics}

\newcommand {\beq} {\begin{equation}}
\newcommand {\eeq} {\end{equation}}
\newcommand {\ber}{\begin{eqnarray*}}
\newcommand {\eer} {\end{eqnarray*}}
\newcommand {\bea}{\begin{eqnarray}}
  \newcommand {\eea} {\end{eqnarray}}
\newcommand{\Nfour} {${\cal N}=4\ $}

\newcommand{\Dslash}{\,{\raise.15ex\hbox{/}\mkern-12mu D}}
\newcommand{\Tr}{{\rm Tr}\,}
\newcommand{\gsim}{\lower.7ex\hbox{$
\;\stackrel{\textstyle>}{\sim}\;$}}
\newcommand{\lsim}{\lower.7ex\hbox{$
\;\stackrel{\textstyle<}{\sim}\;$}}

\def\beqn{\begin{eqnarray}}
\def\eeqn{\end{eqnarray}}

\newcommand{\tr}{\ensuremath{\mathrm{Tr}}}

\def\sumint{\hbox{$\sum$}\!\!\!\!\!\!\int}

\newcommand{\be}{\begin{equation}}
\newcommand{\ee}{\end{equation}}
\newcommand{\ba}{\begin{eqnarray}}
\newcommand{\eps}{\epsilon}
\newcommand{\ea}{\end{eqnarray}}
\newcommand{\bi}{\begin{itemize}}
\newcommand{\ei}{\end{itemize}}

\newcommand{\<}{\langle} 
\renewcommand{\>}{\rangle}  

\newcommand{\la}{\label}
      \def\b{\beta}   \def\g{\gamma}      
\def\d{\delta}      \def\D{\Delta}  \def\e{\varepsilon}
          \def\l{\lambda}     \def\L{\Lambda}
\def\m{\mu}                     
\def\n{\nu}             
\def\r{\sigma}       
\def\t{\tau}          
              
\def\w{\omega}        
\def\HH{{\cal H}}  \def\OO{{\cal O}}

\newcommand{\comma}{\; ,}
\newcommand{\period}{\; .}

\begin{document}
\begin{titlepage}

\vskip 0.5cm

\centerline{{\Large \bf Domain Walls and Metastable Vacua in}}

\vskip 0.2cm

\centerline{{\Large \bf Hot ``Orientifold Field Theories''}}
\vskip 1cm
\centerline{\large Adi Armoni $^{a}$, Chris P. Korthals Altes $^{b}$ and Agostino Patella $^{a}$}

\vskip 0.3cm
\centerline{\it $^{a}$ Department of Physics, Swansea University,}
\centerline{\it Singleton Park, Swansea, SA2 8PP, UK}

\vskip 0.3cm
\centerline{\it $^{b}$ Centre Physique Theorique au CNRS,}
\centerline{\it Case 907, Campus de Luminy,}
\centerline{\it F-13288, Marseille, France}

\vskip 1cm

\begin{abstract}

We consider ``Orientifold field theories'', namely $SU(N)$ gauge theories
 with Dirac fermions in the two-index representation at high temperature. When $N$ is even these theories exhibit a spontaneously broken $Z_2$ centre symmetry.
We study aspects of the domain wall that interpolates between the two vacua
of the theory. In particular we calculate its tension to two-loop order. We compare its tension to the corresponding domain wall in a $SU(N)$ gauge theory
with adjoint fermions and find an agreement at large-$N$, as expected from planar equivalence between the two theories. Moreover, we provide a non-perturbative proof for the coincidence of the tensions at large-$N$.  We also discuss the vacuum structure of the theory when the fermion is given a large mass and argue that there exist $N-2$ metastable vacua. We calculate the lifetime of those vacua in the thin wall approximation. 

\end{abstract}

\end{titlepage}

\section{Introduction}

When a global discrete symmetry is spontaneously broken there exist
 domains walls that interpolate between the various degenerate vacua. An important example is the case of
 the deconfining phase of gauge theories that admit a non-trivial centre symmetry. Due to asymptotic freedom, the high temperature phase is controlled by a weak gauge coupling, hence it is possible to calculate
the domain walls tension order by order in perturbation theory. The tension of the pure $SU(N)$ Yang-Mills theory was
 calculated at the one-loop order in \cite{Bhattacharya:1990hk,Bhattacharya:1992qb}. The calculation was extended to two-loop and three-loop order in \cite{Giovannangeli:2001bh} and \cite{Giovannangeli:2002uv}.
Up to two loop order the tension exhibits a Casimir scaling. A deviation from Casimir scaling 
was found at three-loop order \cite{Giovannangeli:2002uv}. Following \cite{Aharony:1998qu}
the tension of the deconfining domain walls in \Nfour SYM was calculated at both one-loop and strong coupling in \cite{Armoni:2008yp} and extended to two  loop order in \cite{KorthalsAltes:2009dp}.

In this paper we focus on ``orientifold field theories'' (or OrientiQCD), namely on $SU(N)$ gauge theories
with Dirac fermions in either the two-index symmetric or the two-index antisymmetric representations. When $N$ is even the centre symmetry is $Z_2$ \cite{DelDebbio:2008ur}. Therefore there exists a domain wall that interpolate between the two vacua of the theory. The wall tension is ${\cal O}(N^2)$. 

The main motivation for our investigation is the large-$N$ equivalence between $SU(N)$ theory with $N_f$ Dirac fermions in the two-index (either symmetric or antisymmetric) representation and $SU(N)$ theory with $N_f$ Majorana fermions in the adjoint representation~\cite{Armoni:2003gp,Armoni:2004ub}. Planar equivalence states that OrientiQCD and AdjQCD (adjoint QCD) become equivalent in a common sector of bosonic charge conjugation (C-parity) invariant states if and only if C-parity is unbroken in OrientiQCD \cite{Unsal:2006pj}. To present, both analytic and numerical works focused on light states, namely on glueballs and mesons as well as the quark condensate. Heavy objects, such as domain walls, were not yet investigated.

 In the deconfining phase the equivalence between the two theories holds in the common vacua of the the theories, namely when the Polyakov loop expectation value is either $1$ or $-1$ \cite{Unsal:2006pj}. Apriori it is not clear whether the equivalence holds for the domain wall, since it interpolates between different vacua. In addition the domain wall tension is ${\cal O}(N^2)$, namely it is a ``heavy'' object, and it is not obvious that large-$N$ equivalence should hold for such heavy objects.\footnote{${\cal O}(N^2)$ domain walls are described by wrapped NS5 branes in the dual string theory \cite{Armoni:2008yp}.}

We carry out a two-loop calculation of the domain wall tension for an ``orientifold theory'' with $N_f$ massive fermions. The result is that the tension of the wall matches the tension of the corresponding wall of the adjoint fermions theory in the large-$N$ limit, namely that the leading ${\cal O}(N^2)$ contribution is the same in both theories.

In addition to the explicit two-loop calculation we provide a non-perturbative proof for the large-$N$ coincidence of the tensions. The proof is based on the orientifold planar equivalence as formulated in terms of the coherent states~\cite{Yaffe:1981vf,Unsal:2006pj}. The main ingredient will be the reformulation of the domain wall tension at finite temperature as the tension of a Dirac magnetic string at zero temperature but on a compactified space.

Another part of our paper concerns metastable vacua in ``orientifold'' theories. When the two-index fermions are given infinite mass they decouple and we end-up with the Yang-Mills deconfining vacua, namely $N$ degenerate vacua which correspond to the spontaneously broken $Z_N$ centre symmetry of the pure Yang-Mills theory. Therefore at large (but finite) value of fermion mass there should be two real vacua and $N-2$ metastable vacua. We confirm this scenario by an explicit one-loop calculation of the potential for the Polyakov loop. We then discuss the lifetime of the metastable vacua in the thin wall approximation~\cite{Coleman:1980aw,Kobzarev:1974cp}.

The paper is organized as follows: in section 2 we describe the general framework of the paper. In section 3 we present the results of the two-loop
wall tension calculation (leaving the technical details to the appendix). In section 4 we provide the non-perturbative proof for the large-$N$ coincidence of the wall tensions. Section 7 is devoted to the metastable vacua and their lifetime. In section 6 we discuss our results.

\section{Two guises for the 't Hooft loop}\la{sec:guises}

This section provides a platform for the perturbative approach in section \ref{sec:perturbation} and the non-perturbative proof in section \ref{sec:formal}. We will review the original formulation by 't Hooft~\cite{'tHooft:1977hy} for a spatial loop in a theory with non-zero temperature (so fermion fields will be antiperiodic in the Euclidean time direction), in terms of a discontinuous gauge transformation. This one is strictly related to the usual formulation as an electric flux loop operator, which will be useful for the perturbative proof (to low orders) of the equivalence. 

Rotating one of the loop sides in the temporal direction is the starting point for a Hamiltonian formulation, where one studies the propagation of a magnetic flux tube at zero temperature in a 3D space with one compact periodic dimension. The fermion fields  are still antiperiodic in that direction. This Hamiltonian formulation will be useful for the non-perturbative proof of the equivalence in section \ref{sec:formal}.

\subsection{Continuum formulation}

We start with the Hamiltonian formulation on a 3D space (infinite in every direction)\footnote{This subsection relies on the work in ref.~\cite{KorthalsAltes:2000gs}.}. The Hamiltonian  $H$ is 
    built from the electric field strengths $\vec E$ and the magnetic field strengths $\vec B$.  

Apart from the $SU(N)$ gauge field there are  fermionic fields $\psi$, represented by Hermitean $N\times N$ matrices. We are interested only in adjoint and two-index fermions. The theory with adjoint fermions is invariant under the full center $Z_N$, while the theory with symmetric or antisymmetric two-index fermions is invariant under the $Z_2$ subgroup if $N$ is even. The Hamiltonian consists of a gauge part and a fermionic part, minimally coupling fermions to the gauge field through the covariant derivative  $D(A)\psi$ which is $\partial \psi+ig[A,\psi]$ or $\partial \psi+ig(A\psi+\psi A^T)$  (for adjoint and two-index fermions respectively).

Choosing an oriented surface $S$ subtended by the oriented loop $L$, the 't Hooft loop was originally~\cite{'tHooft:1977hy} defined as the unitary operator
\be
V_k(L)=\exp\left\{ \frac{i}{g} \int [ E_b.\vec D(A)_{bc}+g\tr\psi^\dagger T_c \psi] \w_k^c (\vec{x}) d^3x \right\} \ ,
\la{defvk}
\ee
representing a time-independent local gauge transformation $\exp(i T_a \w_k^a(\vec x))$, that has a discontinuity of $z_k=\exp\left(ik{2\pi\over N}\right)$ through the surface $S$. The dot stands for summing over spatial degrees of freedom. This operator depends in general on the particular choice of the surface $S$. However, if $z_k$ belongs to the symmetry subgroup of the theory, the surface can be arbitrarily deformed (provided that its contour is unchanged) via a regular gauge transformation. Therefore in this particular case, the operator $V_k(L)$ acting on physical (i.e. gauge invariant) states depends only on the loop $L$ and not on the particular choice of the surface $S$.

The immediate effect of $V_k(L)$ is that any Wilson loop $W(C)$ in the fundamental representation with $C$ winding once around $L$ in a clockwise direction will pick up a phase $z_k$ ('t Hooft algebra):
\be
V_k(L)W(C)V_k^\dagger(L)=z_kW(C) \ .
\la{thooftalgebra}
\ee

From standard considerations ~\cite{KorthalsAltes:2000gs, KorthalsAltes:2009dp}, up to a regular gauge transformation and provided that $z_k$ belongs to the symmetry subgroup of the theory, the 't Hooft loop can be rewritten as an electric flux operator:
\be
\hat V_k(L) =\exp \left(\frac {4\pi i}{g}\int_S d\vec S.\tr\vec EY_k \right) \ .
\la{efluxrep}
\ee
The numerical $N\times N$ matrix $Y_k$ equals
\be
Y_k={1\over N}\mbox{diag}(\underbrace{k,k,....,k}_{N-k \textrm{ times}},\underbrace{k-N,k-N,....,k-N}_{k \textrm{ times}}) \ ,
\la{defyk}
\ee
generating the center group element $z_k{\bf 1}=\exp(2\pi i Y_k)$.

Deep in the deconfined phase the electric flux is due to almost free statistically independent screened gluons, and elementary arguments show that its thermal average falls off exponentially with the area of the loop~\cite{Giovannangeli:2001bh,KorthalsAltes:2000gs}:
\be \label{eq:thermalev}
\langle V_k(L) \rangle = \frac{1}{Z(T)} \tr [ V_k(L) e^{-H/T} ] \simeq A e^{-\sigma(T) L_xL_y} \ ,
\ee
where $Z(T)=\tr \ e^{-H/T}$ is the partition function, and the trace is taken only on physical (i.e. gauge invariant) states.

\subsection{'t Hooft loop as magnetic correlator}

If we write the thermal expectation value~\eqref{eq:thermalev} in the path integral formulation, the insertion of the operator $V_k(L)$ is implemented as some twisted boundary conditions for the gauge field in the thermal direction:
\be \label{eq:twistedbc}
A_\mu(x,y,z,\t+1/T) = A_\mu(x,y,z,\t) + 2 \pi Y_k \delta_{\mu,z} \chi_{[0,L_x]}(x) \chi_{[0,L_y]}(y) \delta(z) \ .
\ee
Since we are dealing with the Euclidean spacetime, we can now interpret $y$ as the thermal axis. The 3D space is identified by the coorinates $(x,z,\t)$ and the direction $\t$ is a spatial compact direction.  The Hamiltonian in the 3D space is written as $\HH$. The boundary conditions above introduce a twist (localized on a string) in the compact spatial direction, only in the time lapse $[0,L_y]$.

From a Hamiltonian point of view, if $\mathcal{S}_k(L_x)$ is the unitary operator which creates the twist\footnote{This operator produces field configurations which are multivalued on the compact direction.}, then the thermal expectation value of the 't Hooft loop can be written as the zero-temperature Euclidean correlator of the twist operator~\footnote{For simplicity we choose the vacuum to have zero energy.}:
\be \label{eq:magneticcorrelator}
\langle V_k(L) \rangle_T = \langle 0 | \mathcal{S}_k(L_x)^\dag e^{-\HH L_y} \mathcal{S}_k(L_x) | 0 \rangle \ .
\ee
 
When ${\cal S}_k(L_x)$ acts on a small closed loop $\g$ in the $z-\t$ plane that encircles the string identified by $\t=z=0$ and $0 < x < L_x$ at a fixed time, a $z_k$ factor is produced:
\be
{\cal S}_k(L_x)^\dag W(\g){\cal S}_k(L_x)=z_kW(\g) \ .
\la{magneticfluxoperator}
\ee
Notice that this is different from the 't Hooft algebra~\eqref{thooftalgebra}, since the operator ${\cal S}_k(L_x)$ describes a open string-like singularity. Therefore ${\cal S}_k(L_x)$ can be interpreted as the creation operator for a Dirac string carrying a magnetic flux of strength $z_k$, ending in a Dirac monopole-antimonopole pair. This operator causes the Polyakov loop operator
\be
{\cal P}\exp(i\int_0^{1/T} d\t A_\t(x,z,\t))
\ee
as a function of $z$ to jump with the $z_k$ factor when it crosses the location of the Dirac string at $z=0$.  This is based on the operator identity for the ${\cal S}$-transformed Polyakov loops:
\begin{flalign}
& \lim_{\d z\rightarrow 0}{\cal P}\exp(i\int_0^{1/T} d\t A^{{\cal S}}_\t(x,z=-\d z,\t)) \times \nonumber \\
& \qquad \times {\cal P}\exp(-i\int_{1/T}^0 d\t A^{{\cal S}}_\t(x,z=\d z,\t))=z_k \ ,
\la{jump}
\end{flalign}
the left hand side being a closed loop around the Dirac string at $z=\t=0$.

\subsection{Lattice discretized Hamiltonian}

The lattice regularization is particularly useful because it smooths out the singularity in the the boundary conditions~\eqref{eq:twistedbc}. Moreover it provides a manifestly gauge covariant formulation. We remind that in the Hamiltonian formalism only the 3D space is discretized, while the time ($y$ in the notation of the previous subsection) stays a continuous parameter.

Formally we just need to replace the Hamiltonian in eq.~\eqref{eq:magneticcorrelator} with the lattice version ($\mathbf{x}=(x,z,\t)$ are the integer valued coordinates of  three space):
\be \la{latticehamiltonian}
\HH_L={a^2g^2\over 2} \sum_{\mathbf{x}} \tr{ \vec{\cal E}(\mathbf{x})}^2+ {1\over{2 g^2a}} \sum_{\Box}\Re e \tr({\bf 1}-U_{\Box}) +\HH_{F} \ ,
\ee
where the operators $U_{\Box}$ are the Wilson loops around the smallest square paths on the lattice (plaquettes), and $\HH_F$ is the fermionic contribution to the lattice Hamiltonian. We do not need to specify the form of $\HH_F$. We will just assume that it inherits the center symmetry subgroup from its continuum counterpart.

The Hamiltonian $\HH_L$ is transformed by ${\cal S}_k(L_x)$ into the twisted Hamiltonian
\be
\HH_L^t\equiv {\cal S}_k(L_x)^\dag \ \HH_L \  {\cal S}_k(L_x)
\ee
using eq.~\eqref{magneticfluxoperator}. The operator ${\cal S}_k(L_x)$ twists with a factor $z_k$ all the plaquettes that are pierced by the Dirac string:
\ba
\HH^t_L&=&{a^2g^2\over 2} \sum_{\mathbf{x}}\tr{ \vec{\cal E}(\mathbf{x})}^2+
{1\over{2g^2a}} \sum_{\textbf{twisted } \Box} \Re e \tr({\bf 1}-z_kU_{\Box}) \nonumber\\&+&
{1\over{2g^2a}} \sum_{\textbf{nontwisted } \Box}\Re e \tr({\bf 1}-U_{\Box})
+\HH_F.
\la{twistedhamiltonian}
\ea

With this definition the expectation value of the 't Hooft loop becomes:
\be \label{eq:latticemainequation}
\langle V_k(L) \rangle_T = \langle 0 | e^{-\HH_L^t L_y} | 0 \rangle \ .
\ee

This twisted Hamiltonian will be used in section \ref{sec:formal}.  Note that it is Hermitean
for all twists, but charge conjugation invariant only if $z_k=\pm1$. 

\begin{figure}
\begin{center}
\includegraphics[width=11cm,height=2cm]{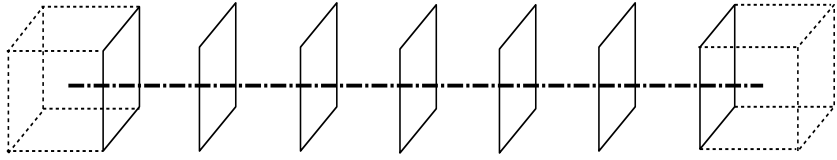}
\end{center}
\caption{\label{fig:monpiccopy}Monopole anti-monopole pair induced by
twisting the z-$\t$ Wilson loops (plaquettes in lattice formulation) pierced by the Dirac string in the $x$ direction. This string 
is propagated in the $y$ direction.} 
\end{figure}

\section{Domain wall equivalence  in perturbation theory}\la{sec:perturbation}

  In this section we discuss the perturbative equivalence of  the two-index $SU(N)$ gauge theory (often called ``orientifold field theory'' or OrientiQCD) with ${\cal N}=1$ $SU(N)$ theory (also called AdjQCD) up to terms of $\OO(1/N)$.
 
 The two-index spinor $\psi^{ij}$ carries two $SU(N)$ indices $i$ and $j$, running from $1$ to $N$. They refer to the transformation property of $\psi^{ij}$ under an $SU(N)$ element $U$ in the fundamental representation:
 \be
 \psi^{ij}\rightarrow (U\psi U^T)^{ij}
\la{transfofermi}
 \ee
\noindent  which leads for even $N$ to invariance under the subgroup $Z_2$ of the full centre group $Z_N$.
 
  We start with the action of the two-index theory. Apart from the gauge field action it contains the minimally coupled fermion action:
 \be
 \overline\psi \g_{\mu}D_{\mu}(A)\psi. 
  \la{fermiaction}
  \ee  
  
  Because of eq.(\ref{transfofermi}) the covariant derivative acts as:
  \be
  D_\mu(A)\psi=\partial_\mu \psi + i A_\mu \psi + i \psi A_\mu^T.
  \ee
 instead of a commutator as behooves an adjoint fermion. The $\g$ matrices are Hermitean.
 
 \subsection{One loop approximation to the tension}

 The effective action $U(P)$ for the Polyakov loop $P(C)$ consists of a kinetic part
 and a potential part:
 \be
 U(P)=K(P)(\partial P)^2+V(P).
 \la{effaction}
 \ee
 In perturbation theory the kinetic term starts with the classical term ($\tilde g^2$ is the 't Hooft coupling):
 \be
 K(P)={1\over{g^2}}(1+\tilde{g}^2K_1(P)+...),
\la{kinform}
 \ee
 \noindent whereas  the potential can be written as:
 \be
 V(P)=V_1(P)+\tilde {g}^2V_2(P)+....
 \ee
 Note the classical kinetic energy is proportional to $1/g^2$, compared to the $\OO(1)$ potential term coming from a fluctuation determinant.. This 
 means the profile $C(z)$ is slowly varying, $\sim gz$.

 We  are therefore allowed to evaluate the effect of the fluctuations around a {\it constant} background value C of 
 the phase of the Polyakov loop:
 \be
 A_\m=C\d_{\m,0}+gQ_\m.
 \ee

  We will take Feynman background gauge.

 The first term in the potential is given by the fluctuation determinant:
 \be
 V_1(P(C))={T\over {\mbox{Vol}}} \Big[{1\over 2}tr_b\log (-\delta_{\m,\n}D(C)^2_-)-Tr_b\log(-D(C)^2_-)-Tr _f\log (i\g.D(C)_+)\Big]
\la{determinant}
 \ee
 
The $\pm$ sign means a commutator or anti-commutator.  The sum over colour and momentum degrees of freedom is written as  $tr$ and is normalised by the three dimensional volume Vol.
 The suffix b (f)
means doing the trace consistent with periodic (anti-periodic) boundary conditions. $Tr$ sums also the spin degrees of freedom. 
 
The $\g$ matrices are taken care of by squaring the Dirac operator and
correcting with a factor $1/2$ in front of the logarithm. The trace over the
Dirac indices gives a factor $4$, so all in all:
\be
Tr _f\log (i\g.D(C)_+)=2 tr_f\log( -D(C)^2_+).
\la{spin}
\ee
 
  The $Z_2$ symmetry leads to an effective potential
 with two degenerate minima, one where the Polyakov loop matrix  takes the value ${\bf 1}$ and one
 where it equals  $-{\bf 1}$. In terms of the phase matrix C :
 \be
 C=0,  \mbox{and} ~C=2\pi T\times{1\over 2}\mbox{diag}(1,1,.......1,-1,-1,.....,-1)\equiv2\pi T Y_{N/2}.
\la{background}
 \ee
 
 There are ${N\over 2}$ entries with $+1$ and an equal number with $-1$.
 So $P(C)=-{\bf 1}$ as promised.
 
 In both theories the tunnelling path is {\it assumed} to be the straight one connecting the two. That it  describes  a local minimum, i.e. a  valley, is easy to establish.  In gluodynamics the straight path hypothesis has been verified for $3$ and $4$ colours to be indeed a global minimum~\footnote{For the covering group of $SO(2n+4)$, having a $Z_4$ centre group, the same hypothesis leads  to inacceptable results}. For any number of colours in gluodynamics it leads to Casimir scaling, which is verified by lattice results~\cite{deForcrand:2005rg}.  

Adding  the two-index fermion deforms the potential. Only the $Z_2$
symmetry remains respected. We have checked  that the straight line interpolation stays a local minimum.  

So along the tunnelling path  we will take C to be of the form:
\be
C(q)=2\pi T q Y_{N/2}.
\la{path}
 \ee    
     
This is a straight path starting with $P(C(q))=0$ for $q=0$ and ending in  $P(C(q))=-{\bf 1}$ for $q=1$.

\subsection{Evaluation of the determinant}

The key element  in eq.(\ref{determinant})  is the covariant derivative. Since 
the background is constant we expand in a plane wave basis. In this plane wave basis the trace is written like a $d-1$ dimensional integral over momenta and a discrete sum over Matsubara frequencies:
\be
\sumint_{~l}\equiv T\sum_{l_0}\int {d\vec l \over{(2\pi)^{d-1}}}\m^{2\e},
\ee
\noindent with $2\e=4-d$. Only from two-loop order on, the $\m$ dependence will show up.

To diagonalise the colour structure we use the Cartan basis with diagonal matrices  $\l_d,~d=1,....,, N-1$ and off-diagonal matrices $\l^{ij}$ with matrix elements $(\l^{ij})_{lm}={1\over{\sqrt{2}}}\d_{il}\d_{jm}$.

The off-diagonal basis elements  are eigenvectors of  any diagonal matrix $C$:
\be
[C,\l^{ij}]_\pm=(C_i\pm C_j)\l^{ij}.
\la{eigenvector}
\ee
In this basis the diagonalisation is trivial:
\be
D_\m(C)_{\pm}Q_\n^{ij}=i(l_\m+(C_i\pm C_j))
\ee

Because of the explicit form for $C(q)$ (eq.(\ref{background}) and (\ref{path})) the potential becomes:
\be
V_1(q)=2\big({N\over 2}\Big)^2\Big[\sumint_{~l} \log((l_0+2\pi Tq)^2+\vec l^2))-2\sumint_{~l}\log(l_0+2\pi T(q+{1\over 2}))^2+\vec l^2)\Big]
\la{oneloopreducible}
\ee

The colour factor in front of the gauge  contribution is  $2(N/2)^2$ because the commutator in eq.(\ref{eigenvector})  is only contributing if the indices $i$ and $j$ are in different sign sectors of $Y_{N/2}=1/2\mbox{diag}(1,1,.....,1,-1,-1,.....,-1)$.
The same colour factor results from the anti-commutator because in that case the non-vanishing contributions come from the same sign sectors. The factor 2 
in front of the fermionic contribution comes from the spin, eq.(\ref{spin}), of our Dirac fermion. Since for fixed colour a Dirac fermion has twice as much degrees of freedom as the gauge boson this factor 2 is to be expected.  This factor 2 is undone for each one
of the irreducible representations, as we will see in the next subsection.

\subsection{Irreducible components of the Dirac fermion and equivalence}

Up till now we computed with a reducible two-index Dirac fermion. We now split the reducible representation $\psi^{ij}$ into an anti-symmetric and a symmetric
representation with respectively ${1\over 2}N(N\pm 1)$ components.
The colour factor $2(N/2)^2$ in front of the reducible fermionic contribution in eq.((\ref{oneloopreducible})  splits into two parts:
 $ {1\over 2}N({N\over 2}\pm 1)$ for symmetric and anti-symmetric representation~\footnote{Note that the $SU(2)$ antisymmetric representation does not contribute, as it is a colour singlet.}. 
 
 For large-$N$ both symmetric and antisymmetric representations contribute
 with the same colour factor  $({N\over 2})^2$.  Combined with the spin factor 2
 both for the symmetric and anti-symmetric representation we have from
 eq.(\ref{oneloopreducible}):
 \be
 V_{1\pm}(q)=2({N\over 2})^2\Big[\sumint_{~l} \log((l_0+2\pi Tq)^2+\vec l^2))-(1\pm{2\over N})\sumint_{~l}\log(l_0+2\pi T(q+{1\over 2}))^2+\vec l^2)\Big]
\la{oneloop}
 \ee 
 
For large-$N$ both the symmetric and antisymmetric representation contribute
 the same colour factor  $({N\over 2})^2$.  The equivalence with the supersymmetric version is then very plausible since omitting the thermal
 boundary condition for the fermion, the boson and fermion contributions for $V_1(q)$ cancel out.
 
Indeed our earlier result~\cite{Armoni:2008yp, KorthalsAltes:2009dp} for the ${\cal N}=1$ evaluated for large-$N$ and 
the special channel $k={N\over 2}$ gives the same result for large-$N$.

\subsection{Potential and kinetic terms to two loop order}
First we discuss the potential. In general potentials as function of the Polyakov loop
involve not only vacuum diagrams but also the renormalisation of the loops. The latter gives rise to insertion diagrams and have been extensively discussed~\cite{KorthalsAltes:1993ca}.

The two-loop diagrams for the potential are given in fig. (\ref{fig:susyonetwolooppot}). The only difference with the calculation 
in reference~\cite{KorthalsAltes:2009dp} is given by the fermionic loops in (b2) and (d). Since the latter
come in linearly and the irreducible representations do not mix, we can 
calculate with the reducible representation and split the result later, as in 
the one loop case above.  Moreover in diagram (d) only the derivative (with respect to the variable $q$)
of the one loop result comes in. The inserted renormalisation of the Polyakov loop is to this order independent of the fermion content. So we know that the equivalence will work for diagram (d).

As there are only four different sum integrals involved 
we list them here:
\ba
\sumint_{~l} \log((l_0+2\pi T q)^2+\vec l^2)&=&\hat B_4(q)\\
\sumint_{~l}  {(l_0+2\pi T q)\over {(l_0+2\pi Tq)^2+\vec l^2}}&=&\hat B_3(q)\\
\sumint_{~l} {1\over{(l_0+2\pi Tq)^2+\vec l^2}}&=&\hat B_2(q)\\
\sumint_{~l} {(l_0+2\pi T q)\over{((l_0+2\pi T q)^2+\vec l^2)^2}}&=&\hat B_1(q).
\la{bernoullihat}
\ea
The functions involved are proportional to the Bernoulli polynomials 
with the suffix indicating their order. They are shown in Appendix \ref{ap:twoloop}.
\begin{figure}
\begin{center}
\includegraphics[width=10cm]{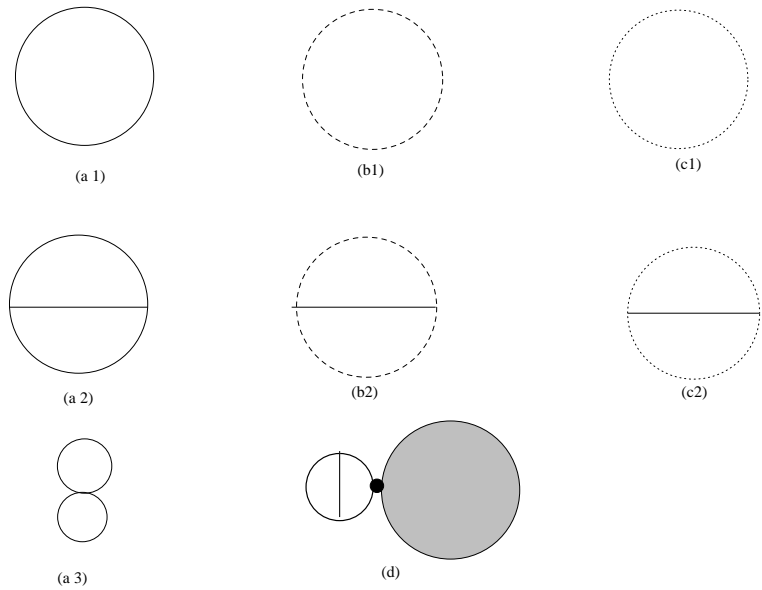}
\end{center}
\caption{One and two loop contributions to the effective potential. Continuous lines are bosons, broken lines are fermions and dotted lines are ghosts. In (a) the bosonic contributions, in (b) the fermionic contributions and (c) the ghost contribution are shown. In (d) 
 the Polyakov loop (fat circle), renormalized by one gluon exchange, is inserted into the sum of the loops (a1), (b1) and (c), given by the shaded blob.}
\la{fig:susyonetwolooppot}
\end{figure}

So the {\it only} diagram to analyse is diagram (b2).
For $k={N\over 2}$  the result is for finite N:
 
\ba
V_{2\pm}&=&{N^2\over 4}\Bigg[\hat B_2(q)^2+2\hat B_2(0)\hat B_2(q) -(1\pm {2\over N})\Big\{-(\hat B_2(q+1/2))^2+2\hat B_2(q)\hat B_2(q+1/2)+\nonumber\\
&+&2\Big(\hat B_2(0)\hat B_2(q+1/2)+\hat B_2(1/2)\hat B_2(q)-\hat B_2(1/2)\hat B_2(q+1/2)\Big)\nonumber\\
&-&{4\over N^2}\Big(\hat B_2(q+1/2)^2-2\hat B_2(0)\hat B_2(q+1/2)\Big)\Big\}\nonumber\\
&+&4\hat B_1(q)\Big(\hat B_3(q)-(1\pm{2\over N})\hat B_3(q+1/2)\Big)\Bigg]
\la{v2finiteN}
\ea
The first two terms and the first term in the last line represent all the purely bosonic two-loop graphs in fig.\ref{fig:susyonetwolooppot},
including the insertion diagram (d). The remaining terms in the curly brackets are the graph (b2) and the very last term is the contribution  from the insertion graph (d).
For $N$ large we compare to the ${\cal N}=1$ result~\cite{KorthalsAltes:2009dp}, using the notation $D_k(q)=B_k(q)-B_k(q+1/2)$ to render the SUSY limit explicit:
\be
 V_2=({N\over 2})^2\bigg[\bigg(\hat D_2(q)+\hat D_2(0)\bigg)^2-\hat D_2(0)^2+4\hat B_1(q)\hat D_3(q)\bigg].
\la{n1susy}
\ee
Indeed the equivalence is valid  for the two loop potential.

We could have seen this result almost directly by  working in 't Hooft's double line notation. For the gluon we write:
 \be
( \d_{ii'}\d_{jj'}-{1\over N}\d_{ij'}\d_{i'j})/l_{ii'}^2.
\la{1overncorrprop}
\ee 

\noindent with $l_{ii'}^2=(l_0+C_i-C_{i'})^2+\vec l^2$. 

Only in the two-index case the arrows on the fermion  propagators run parallel.

The colour shift is identical for adjoint  
or two-index case for colour charge $Y_{N/2}$, eq.(\ref{background}).

The $1/N$ part is the $U(1)$ gluon we have to subtract from the first term. We can then sum over the 
indices as if they were independent as they would be in the $U(N)$ case.  

In the ${\cal N}=1$ case the adjoint fermion does not couple to the $U(1)$ gluon.  Hence, in the large-$N$ limit,  the diagram (b2) in fig. (\ref{fig:susyonetwolooppot}), with  the reducible two-index Dirac fermion and with the adjoint Dirac fermion running through the loop are equal if the colour charge in the vertices is $Y_{N/2}$ as shown in fig.(\ref{fig:doubleline}).  The same equality holds when
as in the ${\cal N}=1$ case a Majorana fermion is used, and in the two-index case an irreducible component appears.

\begin{figure}
\begin{center}
\includegraphics[width=8cm]{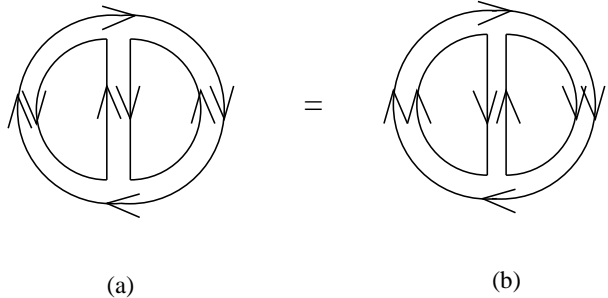}
\end{center}
\caption{Double line representation of the graph in fig. (\ref{fig:susyonetwolooppot}) (b2):  equality in the large-$N$ limit between an adjoint Dirac fermion (a) and a two-index Dirac fermion (b)
running through the fermion loop. The difference  in the (anti-)parallel arrows of the (adjoint) two-index fermions is irrelevant for the charge $Y_{N/2}$ in the vertices. }
\la{fig:doubleline}
\end{figure}

\subsection{Kinetic term}
The remainder of this section is devoted to the one loop kinetic term shown in  fig.(\ref{fig:susyoneloopkin}).   

Clearly the first graph involves the renormalisation of the coupling. 

The beta function coefficients are
\ba
b_b&=&-11/3\nonumber\\
b_{f}&=&{2\over 3}\nonumber\\
\b_0&=&b_b+b_f
\la{betadef}
\ea


\begin{figure}
\begin{center}
\includegraphics[width=8cm]{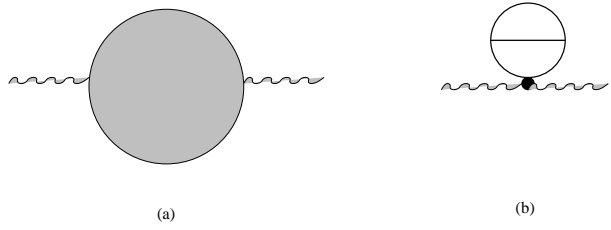}
\end{center}
\caption{The contributions to the one loop effective kinetic term. The wiggly lines are $q$  lines. The  shaded blob in (a) is the sum of one loop boson, fermion and ghost graphs. In panel (b) the renormalization of the Polyakov loop is shown.}
\la{fig:susyoneloopkin}
\end{figure}
In terms of the 
renormalized coupling  $g(T)$ of ref.~\cite{Bhattacharya:1990hk}:
\be
{1\over{\tilde g^2(T)}}={1\over{\tilde g^2}}\bigg\{1+\bigg({\tilde g\over {4\pi}}\bigg)^2b_0\bigg[{1\over \e}+\psi(1/2)+\log\bigg({\m^2\over{\pi T^2}}\bigg)\bigg]\bigg\},
\la{couplingdef}
\ee
\noindent we obtain for the kinetic  term ($\psi(q)$ is the logarithmic derivative of the gamma function):
\ba
K_{1\pm}&=&\left ({N\over 2} \right )^2{(2\pi Tq'(x))^2\over{\tilde g^2}(T)}\Bigg(1-
\bigg({\tilde g(T)\over{4\pi}}\bigg)^2\bigg\{\Big(b_b(\psi(q)+\psi(1-q))\Big)+\nonumber \\
&+&\bigg((1\pm{2\over N})\Big(b_{f}(\psi(q+{1\over 2})+\psi_{reg}(-q+{1\over 2}))+{13\over 3}\Big)
+4\bigg)\bigg\}\Bigg).
 \la{renkinterm}
 \ea
The last term is due to the insertion diagram (b) in fig (\ref{fig:susyoneloopkin}). The $\psi$ function comes in through summation over the Matsubara modes:
\be
\sum_n{1\over{|n+q|^{5-d}}}={1\over {\e}}-(\psi(q)+\psi(1-q)).
\la{psi}
\ee

Let us compare this result for large $N$ to the ${\cal N}=1$ result~\cite{KorthalsAltes:2009dp}:
\ba
K_{1\pm}&=&\left ({N\over 2} \right )^2{(2\pi Tq'(x))^2\over{\tilde g^2}(T)}\bigg(1-\bigg({\tilde g(T)\over{4\pi}}\bigg)^2\bigg\{b_b(\psi(q)+\psi(1-q))+\nonumber \\
&+&b_f((\psi(q+{1\over 2})+\psi_{reg}(-q+{1\over 2}))+{13\over 3}+4\bigg\}\bigg).
 \la{renkintermn1}
 \ea
Also the one loop kinetic term shows equivalence.
Note:
\bi
\item the term $\psi_{reg}(1/2-q)$ in the ${\cal N}=1$ case is defined by subtracting the pole part,
\be
\psi(1/2-q)\rightarrow \psi_{reg}(1/2-q)=\psi(1/2-q)-1/(1/2-q).
\ee
This pole originates in the $n=-1$ Matsubara mode (see eq.(\ref{psi}) with argument $q+1/2$ instead of $q$).
Its subtraction is justified~\cite{KorthalsAltes:2009dp} because our fermion is a Majorana spinor  and   then of the two spin states
with Matsubara frequency $-1$, one is not normalisable, and the other one is normalisable but does not contribute to the energy of the wall~\cite{Jackiw:1975fn}. 
\item The question now is how to regulate this pole in our two-index case, eq.(\ref{renkinterm}).  
 If the fermion is a Dirac fermion, with $n=-1$, then there are four states. There are two finite bound state eigenspinors with zero energy. The other two are non-normalisable.  This means that none of 
 the $n=-1$ fermion states contributes to the kinetic energy term, so that we  can work with $\psi_{reg}(1/2-q)$ in eq.(\ref{renkinterm}).
\item In the bosonic contribution there a is pole as well, at $q=0$, but its origin lies in the thermal infra-red. It does not contribute to the tension, since it is folded with the square root of the 
lowest order potential $B_4(q)$. The latter is $\OO(q)$ near $q=0$.

\ei

We conclude that to two-loop order the effective actions of  the two irreducible two-index fermions
are identical to the ${\cal N}=1$, up to terms of $\OO(1/N)$.

\subsection{A note on perturbative equivalence for ${\cal N}=2,4$ SYM at high $T$}

At zero temperature the equivalence between ${\cal N}=2$ and the two-index fermion theory
becomes tenuous, because of the degeneracy of the ground state in the former.
But at high temperature this degeneracy should vanish, as supersymmetry is now broken.

So we felt justified to describe below the perturbative equivalence between them. In doing so we will also include ${\cal N}=4$.

Consider the action with two-index fermions in $D=6$ or $10$ dimensions and compare it
to ${\cal N}=1$ action in the same dimensionality. The two-index fermion has no constraint
in $6$ dimensions, whereas in $10$ dimensions it obeys a Majorana (or alternatively a chiral) constraint.

Upon dimensional reduction the ${\cal N}=1$ $D=6$ and $D=10$ theories become ${\cal N}=2$ and ${\cal N}=4$ super Yang-Mills theories in $D=4$.

We consider these theories at high temperature and will assume that there is a deconfined 
  groundstate with the scalars in the perturbative ground state.

Perturbation theory gives then identical results for the effective potential $U(P)$. 
The reason is that dimensional reduction does not change the integrals, only the
counting of degrees of freedom. But the change in degrees of freedom turns out to be the same
in the two-index theory as in the corresponding SUSY theory. 
 
 To one loop order the counting of the degrees of freedom due to the extra dimensions is simple.
 For the boson sector the comparison is immediate, for the fermion sector  we get in $\d=6$
 an eight dimensional spinor. Like in the four  dimensional case (see eq.(\ref{oneloop})) there are twice as many fermions as bosonic  degrees of freedom, for a given colour. The same is true
 in ten dimensions, where the spinor is sixteen dimensional. This factor two is absorbed by going to the irreducible two-index fermion representations.
 
 For the one loop kinetic energy the same reasoning holds. And also for the insertion diagram (d) in  fig. (\ref{fig:susyonetwolooppot}), because the renormalisation  of the Polyakov loop stays unchanged in this order (Remember the Polyakov loop
 concerns only one polarization direction, the one in the periodic direction). 
 
In two loop order we need to analyse the diagram  (b2) in fig. (\ref{fig:susyonetwolooppot}). All we need to do is convince ourselves that the
equality in fig. (\ref{fig:doubleline}) still holds in the 6 and 10 dimensional reductions
of the respective theories. All what happens is that the fermion loop is corrected 
by not only a vector particle but also scalar particle exchanged. The only difference is
factors of $1\pm{2\over N}$ in the two-index case.

We conclude that at the perturbative level there is, in addition to the ${\cal N}=1$ case, equivalence between ${\cal N}=2, 4$ SYM and the corresponding ``orientifold'' theories.  Note the equivalence is between the effective potentials,
not only for its minimum, the tension, to which we turn now.

\subsection{Tension to two loop order} 
As discussed in previous sections the tension equals the minimum of the
effective action:
\be
\r_\pm=\left ({N\over 2} \right )^2 {4\pi T\over{\tilde g}}\int_0^1 dq \Big[V_{1\pm}^{1\over 2}+{1\over 2}\tilde g^2\Big({V_{2\pm}\over{V_{1\pm}^{1\over 2}}}+K_{1\pm}V_{1\pm}^{1\over 2}\Big)+O(g^3)\Big]
\la{tension}
\ee
All coefficients under the integral sign are those of eq.(\ref{oneloop}), eq.(\ref{v2finiteN}) and eq.(\ref{renkinterm}), with the factor $({N\over 2})^2$ removed. 

The resulting large-$N$ tension for the (anti-)symmetric two-index fermions is, from eq.(93) in~\cite{KorthalsAltes:2009dp}:
\ba
{\r_\pm(T)\over{({N\over 2})^2T^2}}&=&{4\pi^2\over 15}(9-2\sqrt{3}){T\over {m_D(\d)}}\D\bigg[1-{\tilde g^2\over{(4\pi)^2}}\bigg\{\{(-2.92683 ...)\times b_b(\d)\nonumber\\
&+&(3.27471...)\times b_f(\d)+{13\over 6}\}+ \D\times 5 -2\bigg\}\bigg]
\la{rhoresult}
\ea

with $\D=(\d-2)/2$, $m_D^2(\d)={\D\over 2}\tilde g^2T^2$ and

\ba
b_b(\d)&=&-{11\over 3}+{\d-4\over 6}\\
b_f(\d)&=&{\d-2\over 3}. 
 \la{bfunction}
 \ea

As the ratio of $\L_{\overline{MS}}$ to $T_c$ is not (yet) known we cannot plot the 
tension eq.(\ref{tension}) as a function of $T/T_c$.

\section{Orientifold planar equivalence for the wall tension -- a nonperturbative proof}\la{sec:formal}

Orientifold planar equivalence has been proved in different contexts and under different assumptions. A proof for local observables that assumes the perturbative expansion can be found in~\cite{Armoni:2003gp}. The first nonperturbative proof for the partition function and Wilson loops on the continuum was presented in detail in~\cite{Armoni:2004ub}, in which the authors assumed that the fermionic determinant expansion in the worldline formalism is convergent (this assumption is extensively discussed in~\cite{Armoni:2007rf}). A proof for Wilson loops in the lattice-discretized theory, assuming the strong-coupling and large-mass expansions, was discussed in~\cite{Patella:2005vx}. Finally a nonperturbative proof for C-invariant operators was presented in~\cite{Unsal:2006pj}, assuming that charge conjugation is not spontaneously broken and that the coherent-state construction of~\cite{Yaffe:1981vf} describes correctly the large-$N$ limit of gauge theories.

The setup we have in mind in this section is the one of the coherent states. This is particularly useful when dealing with expectation values of observables expressed in the Hamiltonian formalism. A review of the coherent-state formalism and its application to orientifold planar equivalence is beyond the scope of this paper. We will just summarize the main idea, and we will refer the reader to the literature~\cite{Yaffe:1981vf,Unsal:2006pj} for the details.

The large-$N$ limit of gauge theories is classical, in the sense that quantum fluctuations are suppressed. At every value of $N$ it is possible to identify a particular overcomplete set of states (\textit{coherent states}) in the Hilbert space. In the large-$N$ limit the set of coherent states becomes orthogonal and defines the classical phase space of the $N=\infty$ theory. Roughly speaking, the coherent states have an indetermination of order $1/N^2$, hence the indetermination vanishes in the large-$N$ limit. This construction is analogous to the classical limit ($\hbar \to 0$) for the coherent states of the harmonic oscillator (indetermination of order $\hbar$).
\textit{Classical observables} are operators in the Hilbert space that have the following properties: (a) their matrix elements with respect to the coherent states have a well defined large-$N$ limit, (b) they become diagonal in the basis of the coherent states in the large-$N$ limit. In the $N=\infty$ theory classical observables can be represented as functions on the classical phase space, and the commutator goes into the classical Poisson brackets. Roughly speaking, the classical observables are all the observables which are not sensitive to the quantum correlations between coherent states, and for which the large-$N$ factorization holds. Properly normalized Wilson loops (with possible insertions of electric fields and two-index fermions) and also the Hamiltonian are shown to be classical observables, while the vacuum is a coherent state.

Consider now AdjQCD and OrientiQCD. The common sector is defined as the set of coherent states and classical observables that are charge-conjugation invariant. Assume that charge conjugation is not spontaneously broken in both theories, which means that the vacuum is in the common sector. In this setup, orientifold planar equivalence is stated as follows:
\begin{enumerate}
\item a one-to-one map between the common sectors of AdjQCD and OrientiQCD exists;
\item in the large-$N$ limit the matrix elements of classical observables with respect to coherent states in the common sector are the same in AdjQCD and OrientiQCD;
\item in particular the C-even spectrum of classical observables is the same in AdjQCD and OrientiQCD;
\item since the vacuum is in the common sector, vacuum expectation values of classical observables in the common sector are the same in AdjQCD and OrientiQCD.
\end{enumerate}

As explained in Sect.~\ref{sec:guises}, the domain wall tension is related to the thermal expectation value of the 't Hooft loop ($k=N/2$) in the deconfined phase ($T$ is the temperature):
\begin{gather}
V(L) = \exp \left\{ \frac{4 \pi i}{g} \int \tr ( Y_{N/2} \vec{E} ) d\vec{S} \right\} \ , \\
\langle V(L) \rangle_T = \frac{1}{Z(T)} \Tr V(L) e^{-H/T} \propto e^{-L_xL_y\sigma} \ ,
\end{gather}
where $S$ is a rectangular $L_x \times L_y$ surface lying on the $x-y$ plane at $z = 0$.

One can be tempted to use the arguments above, in order to prove that the thermal expectation value of $V(L)$ (and therefore the domain wall tension) is the same in AdjQCD and OrientiQCD in the large-$N$ limit. However, although the 't Hooft loop is charge-conjugation invariant, it is not a classical observable because it does not have a well defined large-$N$ limit (it is $e^{\mathcal{O}(N^2)}$). Moreover orientifold planar equivalence cannot be straightforwardly exported to \textit{thermal} expectation values. In fact in a thermal expectation value also C-odd states, which are not in the common sector, contribute.

In order to bypass this problem, it is useful to express the 't Hooft loop thermal expectation value, as a zero-temperature magnetic correlator (Sect.~\ref{sec:guises}) on a space with a compact dimension of length $1/T$. We will prefer the lattice discretized formulation, since it avoids the technicality of dealing with discontinuous gauge transformations. We recall here the main formula~\eqref{eq:latticemainequation}:
\begin{equation}
\<V_k(L)\> = \<0|\exp(-L_y\HH^t_L)|0\> \ ,
\end{equation}
where $ |0 \rangle$ is the vacuum of the Hamiltonian $\HH_L$. The Hamiltonians $\HH_L$ and $\HH^t_L$ are defined respectively in eqs.~\eqref{latticehamiltonian} and~\eqref{twistedhamiltonian}.

In the large $L_y$ limit, only the ground state $|t\rangle$ (in the C-even sector) of the twisted Hamiltonian $\HH_L^t$ contributes to the numerator:
\begin{equation}
\langle V_n(L) \rangle \simeq | \langle 0 | t \rangle |^2 e^{-L_y E_t(L_x)  } \ ,
\end{equation}
where $E_t(L_t)$ is the energy of $|t\rangle$ with respect to the Hamiltonian $\HH^t_L$. If also $L_x$ is taken large, then the domain wall tension is recovered from the asymptotic behaviour of $E_t(L_x)$:
\begin{equation} \label{eq:ope_sigma}
E_t(L_x) \simeq \sigma L_x \ .
\end{equation}

The equality of the wall tension in the two theories will follow from the equality of $E_t(L_x)$. Consider the normalized twisted Hamiltonian:
\begin{equation}
h_t = \frac{\HH_L^t}{N^2} \ .
\end{equation}
It can be written as the Hamiltonian in absence of twisting plus a correction:
\begin{equation}
h_t = \frac{\HH_L}{N^2} + {1\over{Ng^2a}} \sum_{\textbf{twisted } \Box} \frac{1}{N} \Re e \tr U_{\Box}
\end{equation}
Separately, the normalized Hamiltonian $\HH_L/N^2$ in absence of magnetic string and the correction (just a sum of normalized plaquettes) are classical observables. Their matrix elements with respect to coherent states are finite in the large-$N$ limit. Therefore $h_t$ is a classical observable itself. Moreover, since only the real part of the plaquettes appears, $h_t$ is also charge-conjugation invariant. Therefore it belongs to the common sector. By orientifold planar equivalence, its lowest eigenvalue $E_t(L_x)/N^2$ in the C-even sector is the same in AdjQCD and OrientiQCD. From Eq.~\eqref{eq:ope_sigma}:
\begin{equation}
\lim_{N \to \infty} \frac{\sigma_{Adj}(\beta)}{N^2} = \lim_{N \to \infty} \frac{\sigma_{Or}(\beta)}{N^2} \ .
\end{equation}

\section{One-loop effective potential and the decay rate of the false vacua}

The one-loop effective potential for the Polyakov loop on thermal $R^3 \times S_1$ and massive fermions in the antisymmetric representation is:
\begin{gather}
V(v) = \frac{2 V_3}{\beta^4 \pi^2} \left\{
- \sum_{i \neq j} \sum_{g>0} \frac{\cos g(v_i-v_j)}{g^4} + 2 N_f \sum_{i<j} \sum_{g>0} \frac{(-1)^g \sigma(g m \beta) \cos g(v_i+v_j)}{g^4}
\right\} \comma \\
\sigma(x) = \frac{x^2}{2}K_2(x) \period
\end{gather}

Choosing the point $v_1=\dots=v_N = \frac{2\pi k}{N}$, the energy density is (up to an additive constant that does not depend on $k$):
\begin{equation}
\epsilon(k) = \frac{V(2\pi k/N)}{V_3} = - \frac{4 N_f}{\beta^4 \pi^2} \frac{N(N-1)}{2} \sum_{g>0} \frac{\sigma(g m \beta) \cos \frac{4 \pi k g}{N}}{g^4} \period
\end{equation}

Since $\sigma(x) \simeq \sqrt{\frac{\pi}{8}} x^{3/2} e^{-x}$ in the $x \to \infty$ limit, then for $m \beta \gg 1$ the only term contributing in the sum is $g=1$ and:
\begin{equation}
\epsilon(k) = \frac{V(2\pi k/N)}{V_3} = - \frac{2 N_f}{\beta^4 \pi^2} \sqrt{\frac{\pi}{2}} \frac{N(N-1)}{2} (m \beta)^{3/2} e^{-m \beta} \cos \frac{4 \pi k}{N} \period
\end{equation}

The energy difference between a metastable state and the stable state is:
\begin{equation}
\Delta \epsilon(k) = \epsilon(k) - \epsilon(0) = \frac{2 N_f}{\beta^4 \pi^2} \sqrt{\frac{\pi}{2}} \frac{N(N-1)}{2} (m \beta)^{3/2} e^{-m \beta} \left( 1 - \cos \frac{4 \pi k}{N} \right) \period
\end{equation}

The large-$N$ limit is taken in the following two cases:
\begin{itemize}
\item $k$ is fixed:
\begin{equation}
\Delta \epsilon(k) = \epsilon(k) - \epsilon(0) = \frac{8 N_f}{\beta^4} \sqrt{\frac{\pi}{2}} (m \beta)^{3/2} e^{-m \beta} k^2 \period
\end{equation}
\item $\phi=\frac{2\pi k}{N}$ is fixed:
\begin{equation}
\Delta \epsilon(\phi) = N^2 \frac{2N_f}{\beta^4 \pi^2} \sqrt{\frac{\pi}{2}} (m \beta)^{3/2} e^{-m \beta} \sin ^2 \phi \period
\end{equation}
\end{itemize}

Consider now the following setup: the space is a $R^2 \times S_1$ and antiperiodic boundary conditions have been chosen for the fermions along the compact direction. The system sits in a    
 false vacuum, labeled by $k$, and then decays to the stable vacuum with $k=0$. The case of $SU(3)$ with sextet fermions was recently discussed in \cite{Machtey:2009wu}.

The decay rate of the $k$ false vacuum is computed in terms of the action of the bounce, which is the solution of the classical Euclidean equations of motion, connecting the false vacuum to the true one in time. In the thin-wall approximation, the bounce looks like a bubble in the Euclidean spacetime. In four dimensions, it is a four-dimensional bubble. When a dimension is compactified with small size $\beta=1/T$, it will look like a three-dimensional bubble of radius $R$.

The action of the bubble is given by the sum of the action of the inner part of the bubble plus the contribution of the wall:
\begin{equation}
S(R) = -\frac{4}{3}\pi R^3 \beta \Delta \epsilon_k + 4\pi R^2 \sigma_k 
\end{equation}
where $\Delta \epsilon_k$ is the energy density difference between the false and true vacua, and $\sigma_k$ is the domain wall tension.

The maximum of the action is reached for:
\begin{gather}
\bar{R} = \frac{2\sigma_k}{\beta \Delta \epsilon_k} \\
\bar{S} = \frac{16}{3} \pi \frac{\sigma_k^3}{\beta^2 \Delta \epsilon_k^2}
\end{gather}

The thin-wall approximation is valid if $\bar{R} \mu \gg 1$ where $\mu$ is the second derivative of the potential in the vacuum. This condition is fulfilled in the large-mass limit.

The general formula for the decay rate of a false vacuum is:
\begin{equation}
\Gamma_k \propto \exp \left\{ - \frac{16}{3} \pi \frac{\sigma_k^3}{\beta^2 \Delta \epsilon_k^2} \right\}
\end{equation}

Consider the orientifold theory with a mass $m$ for the fermions. When the mass becomes infinite, the fermions decouple, the theory becomes $Z_N$ symmetric and $\Gamma=0$. When the mass $m$ is very large but finite, the first nonzero contribution to $\Gamma$ is obtained considering the first nonzero contribution to $\Delta \epsilon_k$ (as computed in the section before):
\begin{equation}
\Delta \epsilon_k = \frac{4 N_f}{\beta^4 \pi^2} \sqrt{\frac{\pi}{2}} \frac{N(N-1)}{2} (m \beta)^{3/2} e^{-m \beta} \sin ^2 \frac{2 \pi k}{N} \comma
\end{equation}
and the domain wall tension $\sigma_k$ of pure Yang-Mills:
\begin{equation}
\sigma_k = k(N-k) \frac{4 \pi^2}{3 \beta^2 \sqrt{3 \lambda}} \period
\end{equation}
The one-loop decay rate in this approximation (and every value of $N$) is:
\begin{equation}
\Gamma_k \propto \exp \left\{ -
\frac{k^3(N-k)^3}{N^2(N-1)^2 \sin ^4 \frac{2 \pi k}{N} } 
\frac{2^{9} \pi^{10}}{3^4 N_f^2 (3 \lambda)^{3/2}}
e^{2m\beta} (m\beta)^{-3}
\right\}
\end{equation}

The large-$N$ limit is taken in the following two cases:
\begin{itemize}
\item $k$ is fixed:
\begin{equation}
\Gamma_k \propto \exp \left\{ -
\frac{N^3}{k}
\frac{2^{5} \pi^{6}}{3^4 N_f^2 (3 \lambda)^{3/2}}
e^{2m\beta} (m\beta)^{-3}
\right\} \label{width1}
\end{equation}
\item $\phi=\frac{2\pi k}{N}$ is fixed:
\begin{equation}
\Gamma_k \propto \exp \left\{ - N^2
\frac{\phi^3(2\pi-\phi)^3} {\sin ^4 \phi } 
\frac{2^{3} \pi^{6}}{3^4 N_f^2 (3 \lambda)^{3/2}}
e^{2m\beta} (m\beta)^{-3}
\right\} \label{width2}
\end{equation}
\end{itemize}

\section{Discussion}

In this paper we discussed theories with matter in the adjoint and two-index representations at high temperature. We focused on domain walls.

The main results of our paper are: (i). a two-loop calculation of the tension of the domain wall the interpolate between the vacua with $\langle P \rangle =1$  and $\langle P \rangle =-1$, (ii). a comparison with the tension of the corresponding domain wall in a theory with adjoint fermions, (iii). a proof that at large-$N$ there is an exact equivalence between the domain walls of the two theories and (iv). a calculation of the decay rates of the false vacua in the ``orientifold field theory''.

Our results suggest that planar equivalence holds not only in a given vacuum, but also for objects that interpolate between distinct vacua. Moreover, planar equivalence holds not only for light (namely ${\cal O}(1)$) objects such as glueballs, but also for heavy (in the present case  ${\cal O}(N^2)$) objects such as domain walls.

Concerning the vacuum structure of the large-$N$ and large fermion mass orientifold theories: we learnt that those theories admit $N-2$ false vacua (that become true vacua when the fermion mass becomes infinite and decouples). Those vacua have a very narrow width in the large-$N$ limit: they decay exponentially either as $\exp -N^3$ \eqref{width1} or $\exp -N^2$ \eqref{width2}, depending on the way that the large-$N$ limit is taken.

There are several interesting future directions to explore. It will be interesting to study other theories with a $Z_2$ centre, such as theories based on an orthogonal gauge group and in particular to study their gravity dual. Such theories contain unoriented strings.

It is also interesting to study, from the string dual side the existence and decay of the false vacua into the true vacua of the theory.

Finally, as so little is known about deconfining domain walls from lattice simulation, it will be interesting to carry out simulations of domain walls of theories with fermions in higher representations on the lattice. \\   
 
{\it \bf Acknowledgements.} We thank J. Ridgway for a collaboration in early stages of this work. A.A. wishes to thank the particle physics group at the Weizmann institute for the kind and warm hospitality, where part of this work has been done. A.A. also thanks the ``Feinberg Foundation Visiting Faculty Program'' at the Weizmann Institute of Science.

\appendix
\section{Details about the two loop calculation}\la{ap:twoloop}
 
 The Bernoulli polynomials~\cite{ryzhik} are related to the sum integrals in
 eq.(\ref{bernoullihat}) as:
 \ba
\hat B_{d=4}(x)&=&{2\over 3}\pi^2T^4 B_4(x)\\
\hat B_3(x)&=&{2\over 3}\pi T^3 B_3(x)\\
\hat B_2(x)&=&{1\over 2}T^2 B_2(x)\\
\hat B_1(x)&=&-{T\over{4\pi}} B_1(x)
\la{bernoulli2}
\ea
\noindent and finally ($\eps(x)$ is the sign function):
\ba
B_4(x)&=&x^2(1-|x|)^2\\
B_3(x)&=&x^3-{3\over 2}\eps(x)x^2+{1\over 2}x\\
B_2(x)&=&x^2-\eps(x)x+{1\over 6}\\
B_1(x)&=&x-{1\over 2}\eps(x)
\la{bernoulli3}
\ea

With this definition the relations (\ref{bernoulli2}) are valid in the interval 
$-1\le x\le 1$.

Even (odd) polynomials are even (odd) in $x$. From the original definition (\ref{bernoullihat}) all the $\hat B(x)$ are periodic mod $1$.

We now turn to an analysis of the two-loop formulae for the potential $V_{2\pm}$.
  
 \begin{table}[htdp]
\caption{Counting of the multiplicity of the two-loop planar diagram. $[k]$ is the $Y_k$ sector with $k$ entries $k-N$, and $[N-k]$ is the sector with $N-k$ entries $k$.  $i$,$j$ and $m$ are the indices of the index cycles discussed in the text.The function B in column 4  equals $\hat B_2(x)=\sumint_{~~l} ~((l_0+2\pi Tx )^2+\vec l^2)^{-1}$, as appears in eq.(\ref{bernoullihat}). The full result for the diagram b2 results from multiplying the multiplicities and the weights for every row. 
}
\begin{center}
\begin{tabular}{|c|c|c|c|}
\hline
$[k]$ & $[N-k]$ & $multiplicity$ & $weight$\\
\hline
\hline
$mj $ & $i$ & $k^2(N-k)$ & $B(2(r-1)q+{1\over 2})B({1\over 2})-B(q)(B(2(r-1)q+{1\over 2})+B({1\over 2}))$\\
$i$ & $mj$& $k(N-k)^2$& $B(2rq+{1\over 2})B({1\over 2})-B(q)(B(2rq+{1\over 2})+B({1\over 2}))$\\
\hline
$j$ & $mi$& $k(N-k)^2$& $B(2rq+{1\over 2})B({1\over 2})-B(q)(B(2rq+{1\over 2})+B({1\over 2}))$\\
$mi$ & $j$& $k^2(N-k)$& $B(2(r-1)q+{1\over 2})B({1\over 2})-B(q)(B(2(r-1)+{1\over 2})+B({1\over 2}))$\\
\hline
$m$ & $ij$& $k(N-k)^2$& $B(2r-1q)+{1\over 2})^2-2B(0)B(2r-1q)+{1\over 2})$\\
$ij$ &$m$& $k^2(N-k)$& $B(2r-1)q+{1\over 2})^2-2B(0)B(2r-1)q+{1\over 2})$\\
\hline
$ijm$ &${}$ & $k^3$& $B(2(r-1)q+{1\over 2})^2-2B(0)B(2(r-1)q+{1\over 2})$\\
${}$&$ijm$& $(N-k)^3$& $B(2rq+{1\over 2})^2-2B(0)B(2rq+{1\over 2})$\\
\hline
\end{tabular}
\end{center}
\label{table2loopq}
\end{table}%

If the weights are all the same, then the total multiplicity is $N^3$ by adding up the entries in the multiplicity
column.
 
 We took the indices $i,j,m$ to be independent. Hence we have to 
 correct by subtracting the diagram with the $U(1)$ gluon which has multiplicity $N$ so the total multiplicity at $q=0$ is $N(N^2-1)$.
%
 The weight of the $U(1)$ gluon contribution combined with the multiplicity is:
 \ba
&-&{1\over N}2\Big[k^2\big( (B(2(r-1)q+{1\over 2})^2-2B(0)B(2(r-1)q+{1\over 2})\Big)\nonumber\\&+&(N-k)^2\Big((B(2rq+{1\over 2})^2-2B(0)B(2rq+{1\over 2})\Big)\nonumber\\
&+&2k(N-k)\Big(B((2r-1)q+{1\over 2})^2-2B(0)B((2r-1)q+{1\over 2}))\Big)\Big].
\la{u1}
\ea
 
 The factor ${-1\over N}$ comes from the U(1) part of the double line gluon propagator:

Its contribution is $\OO(1/N^2)$ smaller than the leading term, because of the explicit $1/N$ factor in the
propagator and one index loop less in the diagram.

\bibliography{walls}

\end{document}